\documentclass{appolb}%
\usepackage{amssymb}
\usepackage{epsfig}
\usepackage{amsmath}
\usepackage{amsfonts}
\usepackage{graphicx}%
\begin{document}

\title{Hadronic corrections to muon anomalous magnetic moment within the instanton
liquid model \thanks{Presented at XLV Cracow School of Theoretical Physics, Zakopane, Poland
 June 3- 12, 2005.}}
\author{Alexander E. Dorokhov
\address{Bogoliubov Laboratory of Theoretical Physics, Joint Institute for Nuclear
Research, Dubna, 141980 Russia } }
\maketitle

\begin{abstract}
The current status of the muon anomalous magnetic moment problem is briefly
presented. The corrections to muon anomaly coming from the effects of hadronic
vacuum polarization, $Z^{\ast}\gamma\gamma^{\ast}$ effective vertex and
light-by-light scattering are estimated within the instanton model of QCD vacuum.

\end{abstract}



\section{Muon AMM: experiment vs theory.}

The study of anomalous magnetic moments (AMM) of leptons, $a=(g_{S}-2)/2$,
have played an important role in the development of the standard model (SM).
At present accuracy the electron AMM due to small electron mass is sensitive
only to quantum electrodynamic (QED) contributions. The theoretical error
\cite{KinNio04} is dominated by the uncertainty in the input value of the QED
coupling $\alpha\equiv e^{2}/(4\pi)$. Thus, the electron AMM provides the best
observable for determining the fine coupling constant%
\begin{equation}
\alpha^{-1}=137.035~998~83(51).
\end{equation}

\begin{figure}[h]
\begin{center}
\includegraphics[height=4cm]{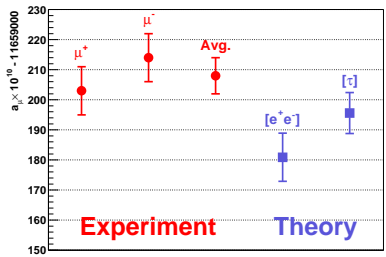}
\end{center}
\caption{Measurements of $a_{\mu}$ by E821 (g-2) Collaboration with the SM
predictions \cite{g-2Coll}. (Theoretical point based on usage of $e^{+}%
e^{-}\rightarrow$hadrons annihilation data is raised up after recent analysis
by SND collaboration \cite{SND}).}%
\label{g2}%
\end{figure}

Compared to the electron, the muon AMM has a relative sensitivity to heavier
mass scales which is typically proportional to $\left(  m_{\mu}/m_{e}\right)
^{2}$.\footnote{The $\tau$-lepton AMM due to $\tau$'s highest mass is the best
for searching for manifestation of effects beyond SM, however, $\tau$-lepton
is short living particle, so it is not easy to make experiment with good
enough accuracy.} At present level of accuracy, the muon AMM gives an
experimental sensitivity to virtual $W$ and $Z$ gauge bosons as well as a
potential sensitivity to other, as yet unobserved, particles in the few
hundred GeV$/c^{2}$ mass range. The muon AMM is known to an unprecedented
accuracy of order of $1$ ppm. The latest result from the measurements of the
Muon $(g-2)$ collaboration at Brookhaven is \cite{g-2Coll}
\begin{equation}
a_{\mu}^{\mathrm{Exp}}\equiv{\frac{1}{2}}(g_{\mu}-2)=11\ 659\ 208\left(
6\right)  \cdot10^{-10}, \label{AMMg-2}%
\end{equation}
which is the average of the measurements of the AMM for the positively and
negatively charged muons (Fig. \ref{g2}). In future, one expects to achieve
more than a factor of $2$ reduction in $a_{\mu}$ uncertainty in planning BNL
E969 experiment \cite{e969} and even more precise g-2 experiment is discussed
in J-PARC with the proposal to reach a precision below $0.1$ ppm \cite{JPARC}.

\begin{figure}[h]
\begin{center}
\includegraphics[height=6cm]{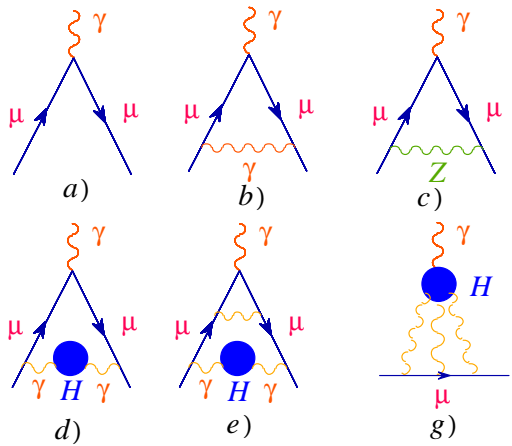}
\end{center}
\caption{The standard model contributions to the muon anomalous magnetic
moment.}%
\label{SM}%
\end{figure}

The standard model prediction for $a_{\mu}$ consists of quantum
electrodynamics, weak and hadronic contributions (schematically presented in
Fig. \ref{SM}). The QED and weak contributions to $a_{\mu}$ have been
calculated with great accuracy \cite{KinNio04}%
\begin{equation}
a_{\mu}^{\mathrm{QED}}=11~658~471.935(0.203)\cdot10^{-10} \label{AMMqed}%
\end{equation}
and \cite{CzMV03}
\begin{equation}
a_{\mu}^{\mathrm{EW}}=15.4(0.3)\cdot10^{-10}. \label{AMMweak}%
\end{equation}

The uncertainties of the SM value for $a_{\mu}$ (Fig. \ref{g2}) are dominated
by the uncertainties of the hadronic contributions, $a_{\mu}^{\mathrm{Strong}%
},$ since their evaluation involve quantum chromodynamics (QCD) at
long-distances for which perturbation theory cannot be employed. Under
assumption that at reached scales there are no New Physics effects one may
estimate the hadronic part of the muon AMM by subtracting the QED and EW
contributions from the experimental result (\ref{AMMg-2})%
\begin{equation}
a_{\mu}^{\mathrm{Strong(Exp)}}=721(6)\cdot10^{-10}. \label{AMMstrong}%
\end{equation}

Below we discuss with some details theoretical status of hadronic
contributions. First, we discuss the phenomenological estimates of the leading
of order $\alpha^{2}$ (LO) hadronic corrections based on usage of inclusive
$e^{+}e^{-}\rightarrow$hadrons and hadronic $\tau$ decays data. Then, one
evaluate the hadronic corrections of leading and next-to-leading (NLO) order
to muon AMM within the instanton liquid model of QCD vacuum (ILM).

\section{Phenomenological estimates of the LO hadronic contributions to muon
AMM}

The LO contribution to the muon AMM comes from the hadronic vacuum
polarization (Fig. \ref{SM}d) and NLO corrections consisting of contributions
which are the iteration of the LO term (Fig. \ref{SM}e) plus the independent
contribution from the light-by-light scattering process (Fig. \ref{SM}g). In
absolute value the LO and NLO terms differ by one order of magnitude, but the
theoretical accuracy of their extraction is comparable and dominates the
overall theoretical error of the SM calculations. All hadronic contributions
are sensitive to the low energy physics and there are no rigorous theoretical
methods based on first principles for the calculations. Thus, to confront
usefully theory with the experiment requires a better determination of the
hadronic contributions.

The LO correction to muon AMM, $a_{\mu}^{\mathrm{hvp}~\left(  1\right)  },$ is
due to the hadronic photon vacuum polarization effect in the internal photon
propagator of the one-loop diagram (Fig.~\ref{SM}d). Using analyticity and
unitarity (the optical theorem) $a_{\mu}^{\mathrm{hvp}~\left(  1\right)  }$
can be expressed as the spectral representation integral \cite{BM61,LBdRGdR}
\begin{equation}
a_{\mu}^{\mathrm{hvp}\left(  1\right)  }=\left(  \frac{\alpha}{\pi}\right)
^{2}\int_{0}^{\infty}dt\frac{1}{t}K(t)\rho_{\mathrm{V}}^{(\mathrm{H})}\left(
t\right)  \ , \label{Amm_rho}%
\end{equation}
which is a convolution of the hadronic spectral function $\rho_{V}%
^{\mathrm{(H)}}\left(  t\right)  $ with the known QED kinematical factor%
\begin{equation}
K(t)=\int_{0}^{1}dx{\frac{x^{2}(1-x)}{x^{2}+(1-x)t/m_{\mu}^{2}}}\ ,
\label{Kfac}%
\end{equation}
where $m_{\mu}$ is the muon mass. The QED factor is sharply peaked at low $t$
and decreases monotonically with increasing $t$. Thus, the integral defining
$a_{\mu}^{\mathrm{hvp}\left(  1\right)  }$ is sensitive to the details of the
spectral function $\rho_{V}^{\mathrm{(H)}}\left(  t\right)  $ at low invariant
masses.\begin{figure}[h]
\begin{center}
\includegraphics[height=6cm]{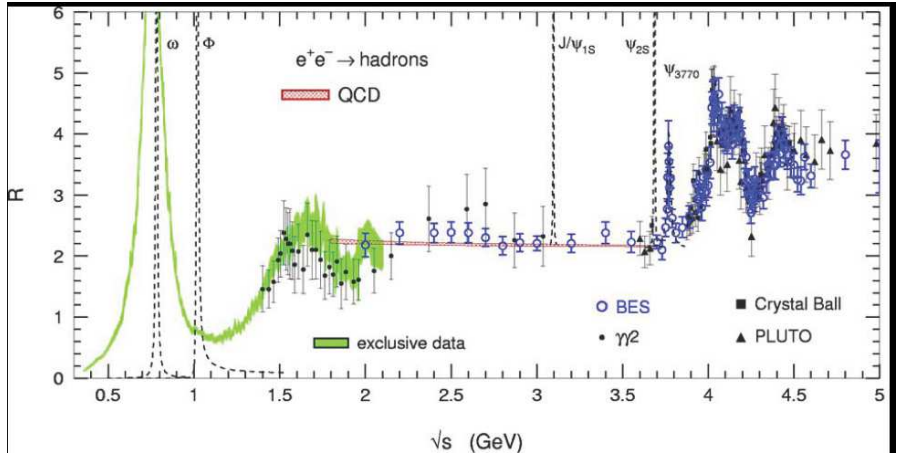}
\end{center}
\caption{Spectral density as measured from $e^{+}e^{-}\rightarrow$hadrons
annihilation, $R(s)=\sigma^{e^{+}e^{-}\rightarrow\mathrm{hadrons}}%
(s)/\sigma^{e^{+}e^{-}\rightarrow\mu^{+}\mu^{-}}(s)$.}%
\label{EEdat}%
\end{figure}

\begin{figure}[h]
\begin{center}
\includegraphics[height=6cm]{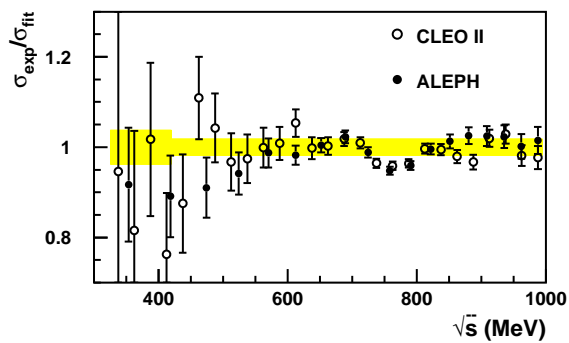}
\end{center}
\caption{The ratio of the $e^{+}e^{-}\rightarrow\pi^{+}\pi^{-}$ cross section
calculated from the $\tau^{-}\rightarrow\pi^{-}\pi^{0}\nu_{\tau}$ decay
spectral function measured in \cite{CLEO,ALEPH2} to the isovector part of the
$e^{+}e^{-}\rightarrow\pi^{+}\pi^{-}$ cross section measured by SND (from
\cite{SND}). The shaded area shows the joint systematic error.}%
\label{SNDfig}%
\end{figure}

At present there is no direct theoretical tools that allow to calculate the
spectral function with required accuracy. Fortunately, $\rho_{V}%
^{\mathrm{(H)}}\left(  t\right)  $ is related to the total $e^{+}%
e^{-}\rightarrow\gamma^{\ast}\rightarrow$ hadrons cross-section $\sigma(t)$ by
$(m_{e}\rightarrow0)$
\begin{equation}
\sigma^{e^{+}e^{-}\rightarrow\mathrm{hadrons}}(t)=4\pi\alpha^{2}\frac{1}%
{t}\rho_{\mathrm{V}}^{(\mathrm{H})}\left(  t\right)  , \label{Sigma_rho}%
\end{equation}
and this fact is normally used to get quite accurate estimate of $a_{\mu
}^{\mathrm{hvp}\left(  1\right)  }$. The condensed form accumulating the data
of different experiments on the hadronic $e^{+}e^{-}$ annihilation is
presented in Fig. \ref{EEdat}.\ Moreover, high precision inclusive hadronic
$\tau$ decay data \cite{ALEPH2,OPAL,CLEO} are used in order to improve the
determination of $a_{\mu}^{\mathrm{hvp}\left(  1\right)  }$. This is possible,
since the vector current conservation law relates the $I=1$ part of the
electromagnetic spectral function to the charged current vector spectral
function measured in $\tau\rightarrow\nu$ +non-strange hadrons. At present, it
is found consistence within the experimental errors between $e^{+}e^{-}$ and
$\tau$ data \cite{SND} (see Fig. \ref{SNDfig}). All these allows to reach
during the last decade a substantial improvement in the accuracy of the
contribution from the hadronic vacuum polarization.

About 91\% of $a_{\mu}^{\mathrm{hvp}\left(  1\right)  }$ comes from $t<(1.8$
GeV$)^{2}$, while 73\% of the corresponding integral is covered by final
$2\pi$ state. The most recent estimates of the dispersion integral for the
$2\pi$-channel in the energy range $0.39<t_{\pi}<0.97$ GeV$^{2}$ which are
based on the $e^{+}e^{-}$ experimental results are following%
\begin{align}
a_{\mu}^{\mathrm{\pi\pi}}  &  =(378.6\pm5.0)\cdot10^{-10}\quad\mathrm{CMD2}%
\text{ }(2003)\cite{CMD2},\nonumber\\
a_{\mu}^{\mathrm{\pi\pi}}  &  =(375.6\pm5.7)\cdot10^{-10}\quad\mathrm{KLOE}%
\text{ }(2004)\cite{KLOE},\nonumber\\
a_{\mu}^{\mathrm{\pi\pi}}  &  =(385.6\pm5.2)\cdot10^{-10}\quad\mathrm{SND}%
\text{ }(2005)\cite{SND}. \label{amupi}%
\end{align}
The contributions of hadronic vacuum polarization at order $\alpha^{2}$ quoted
in the theoretical articles on the subject are given in the Table 1. However,
these analysis do not take into account recent SND\ data which alone may
increase the estimates based on $e^{+}e^{-}$ annihilation by approximately
$(7\div10)\cdot10^{-10}$ (see (\ref{amupi})) making $e^{+}e^{-}$ and $\tau$
data analysis more consistent from one side and more close to experimental
result from other one.

Table 1.\\[0.1cm]Phenomenological estimates and references for the leading
order hadronic photon vacuum polarization contribution to the muon anomalous
magnetic moment based on $e^{+}e^{-}$ and $\tau$ data sets.

\begin{center}%
\begin{tabular}
[c]{|c|c|c|c|c|c|}\hline
& $e^{+}e^{-}\cite{DEHZh03}$ & $\tau\cite{DEHZh03}$ & $e^{+}e^{-}\cite{J}$ &
$e^{+}e^{-}\cite{TrocYnd}$ & $\tau\cite{TrocYnd}$\\\hline
$a_{\mu}^{\mathrm{hvp}~\left(  1\right)  }\cdot10^{10}$ & $696.3\pm9.8$ &
$711.0\pm8.6$ & $694.8\pm8.6$ & $693.5\pm9.0$ & $701.8\pm8.9$\\\hline
\end{tabular}

\end{center}

The higher order hadronic corrections to $a_{\mu}$ are schematically presented
in Figs. \ref{SM} e and g. These diagrams, like leading order contribution,
cannot be calculated in perturbative QCD, but part of them may be estimated
with help of experimental data on inclusive hadronic $e^{+}e^{-}$ annihilation
and $\tau$ decays as \cite{Kra97}\footnote{The second order kernel
$K^{(2)}(t)$ has been evaluated in analytical form in \cite{Barbieri}. For new
formulation of the problem of vacuum polarization effects in higher order
contributions to $(g-2)$ see \cite{Kuraev05}.}
\begin{equation}
a_{\mu}^{\mathrm{hvp}\left(  2\right)  }=-10.1(0.6)\cdot10^{-10}. \label{amu2}%
\end{equation}
This, however, not the case for the so called light-by-light contribution,
$a_{\mu}^{\mathrm{h.~L\times L}}$, (Fig. \ref{SM}g) where one needs to explore
the QCD motivated approaches. The latter has been estimated recently using the
vector meson dominance model supplemented by perturbative QCD\ constraints
\cite{MelnVain03}%
\begin{equation}
a_{\mu}^{\mathrm{h.~L\times L}}=13.6(2.5)\cdot10^{-10}. \label{ammLL}%
\end{equation}

The agreement between the SM predictions and the present experimental values
is rather good. There is certain inconsistencies in use of different sets of
experimental data based on the $e^{+}e^{-}$ and $\tau$ processes in
evaluations of the LO hadronic contribution to the muon AMM. The analysis
based on the $\tau$ decay data and recent $e^{+}e^{-}$ data from SND
collaboration \cite{SND}\ provide the SM results which are in good agreement
with the experimental one. The results based on the $e^{+}e^{-}$ data
published by the CMD \cite{CMD2} and KLOE \cite{KLOE} data support bigger
difference between SM prediction and (g-2) Collaboration result.
Theoretically, the $\tau$ decay data is found \cite{Maltman05} to be more
compatible with expectations based on high-scale $\alpha_{s}(M_{Z})$
determinations; the electroproduction data (CMD, KLOE), in contrast, requires
significantly lower $\alpha_{s}(M_{Z})$. The results favor determinations of
the leading order hadronic contribution to $a_{\mu}$ which incorporate
hadronic $\tau$ decay data over those employing electroproduction data only,
and hence suggest a reduced discrepancy between the SM prediction and the
current experimental value of $a_{\mu}$.

\section{\noindent The Adler function and $a_{\mu}^{\mathrm{hvp}\left(
1\right)  }.$}

\begin{figure}[h]
\begin{center}
\includegraphics[height=6cm]{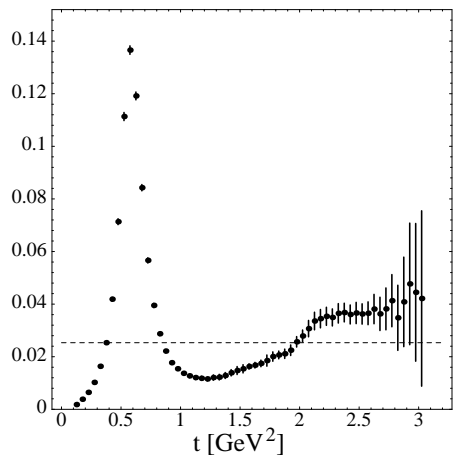}
\end{center}
\caption{The isovector vector spectral function ($1/4\pi^{2})\rho_{V}(t)$ from
hadronic $\tau$- decays \cite{ALEPH2}. The dashed line is the asymptotic
freedom prediction, ($1/4\pi^{2}).$}%
\label{FAleph}%
\end{figure}

Recently, the isovector vector ($V)$ and axial-vector ($A)$ spectral functions
have been determined separately with high precision by the ALEPH \cite{ALEPH2}
and OPAL \cite{OPAL} collaborations from the inclusive hadronic $\tau$-lepton
decays ($\tau\rightarrow\nu_{\tau}+$ hadrons) in the interval of invariant
masses up to the $\tau$ mass, $0\leq s\leq m_{\tau}^{2}$. The vector spectral
function measured by ALEPH is shown in Fig. \ref{FAleph}. It is important to
note that the experimental separation of the $V$ and $A$ spectral functions
allows us to test accurately the saturation of the chiral sum rules of
Weinberg-type in the measured interval. On the other hand, at large $s$ the
correlators can be confronted with perturbative QCD (pQCD) thanks to
sufficiently large value of the $\tau$ mass.

Model estimates of the light quark strong sector of the standard model will be
discussed in the chiral limit, when the masses of $u$, $d$, $s$ light quarks
are set to zero. In this approximation, the $V$ and non-singlet $A$
current-current correlation functions in the momentum space (with
$-q^{2}\equiv Q^{2}\geq0$) are defined as
\begin{align}
\Pi_{\mu\nu}^{J}(q)  &  =i\int d^{4}x~e^{iqx}\Pi_{\mu\nu}^{J,ab}(x)=\,\left[
\left(  q_{\mu}q_{\nu}-g_{\mu\nu}q^{2}\right)  \Pi_{J}(Q^{2})\right]
,\label{PA}\\
\qquad\Pi_{\mu\nu}^{J}(x)  &  =\langle0\left\vert T\left\{  J_{\mu}(x)J_{\nu
}(0)^{\dagger}\right\}  \right\vert 0\rangle, \nonumber\label{Pimn}%
\end{align}
where in the local theory the QCD $V$ and $A$ currents for light
quarks are defined as
\begin{equation}
J_{\mu}=\bar{q}\,\gamma_{\mu}Vq,\qquad J_{\lambda}^{5}=\bar{q}\,\gamma
_{\lambda}\gamma_{5}Aq\,, \label{JAV}%
\end{equation}
the quark field $q_{f}^{i}$ has color ($i$) and flavor ($f$) indices,
$A^{\left(  3\right)  }=\tau_{3}$ is the isospin
matrix of the axial current, and $V=\frac{1}{2}\left(  \frac{1}{3}+\tau
_{3}\right)  $ is the charge matrix. The momentum-space two-point
correlation functions obey (suitably subtracted) dispersion relations,
\begin{equation}
\Pi_{J}(Q^{2})=\int_{0}^{\infty}\frac{ds}{s+Q^{2}}\frac{1}{\pi}\mathrm{Im}%
\Pi_{J}(s), \label{Peuclid}%
\end{equation}
where the imaginary parts of the correlators determine the spectral functions%
\begin{equation}
\rho_{J}(s)=4\pi\mathrm{\operatorname{Im}}\Pi_{J}(s+i0).
\end{equation}
Instead of the polarization function it is more convenient to work with the
Adler function defined as
\begin{equation}
D_{J}(Q^{2}){=-Q}^{2}\frac{d\Pi_{J}(Q^{2})}{dQ^{2}}{\,=}\frac{1}{4\pi^{2}}%
\int_{0}^{\infty}{dt}\frac{Q^{2}}{(t+Q^{2})^{2}}{\,\rho_{J}(t)\,.}
\label{AdlerV}%
\end{equation}

Then, it is possible to express $a_{\mu}^{\mathrm{hvp}\left(  1\right)  }$
given by (\ref{Amm_rho}) in terms of the Adler function by using the integral
representation \cite{6}
\begin{equation}
a_{\mu}^{\mathrm{hvp}\left(  1\right)  }=\frac{4}{3}\alpha^{2}\int_{0}%
^{1}dx\frac{\left(  1-x\right)  \left(  2-x\right)  }{x}D_{V}\left(
\frac{x^{2}}{1-x}m_{\mu}^{2}\right)  , \label{aAd}%
\end{equation}
where the charge factor $\sum Q_{i}^{2}=2/3$, $i=u,d,s,$ is taken into
account. The bulk of the integral in (\ref{aAd}) is governed by the low energy
behavior of the Adler function $D_{V}(Q^{2})$.

The behaviour of the correlators at low and high momenta is constrained by
QCD. In the regime of large momenta the Adler function is dominated by pQCD
contribution supplemented by small power corrections
\begin{equation}
D_{V}(Q^{2}\rightarrow\infty)=D_{V}^{\mathrm{pQCD}}(Q^{2})-\frac{\alpha_{s}%
}{4\pi^{3}}\frac{\lambda^{2}}{Q^{2}}+\frac{1}{6}\frac{\alpha_{s}}{\pi}%
\frac{\left\langle \left(  G_{\mu\nu}^{a}\right)  ^{2}\right\rangle }{Q^{4}%
}+\frac{O_{D}^{6}}{Q^{6}}+\mathcal{O}(\frac{1}{Q^{8}}), \label{Dope}%
\end{equation}
where the pQCD contribution with three-loop accuracy is given in the chiral
limit in $\overline{\mathrm{MS}}$ renormalization scheme by
\cite{3Loop,Chet3}
\begin{equation}
D_{V}^{\mathrm{pQCD}}(Q^{2};\mu^{2})={\frac{1}{4\pi^{2}}}\left\{
1+\frac{\alpha_{s}\left(  \mu^{2}\right)  }{\pi}+\left[  F_{2}-\beta_{0}%
\ln{\frac{Q^{2}}{\mu^{2}}}\right]  \left(  \frac{\alpha_{s}(\mu^{2})}{\pi
}\right)  ^{2}\right.  + \label{DpQCD}%
\end{equation}%
\[
{+}\left.  \left[  F_{3}-\left(  {2}F_{2}\beta_{0}{+\beta_{1}}\right)
{\ln{\frac{Q^{2}}{\mu^{2}}}+\beta_{0}^{2}}\left(  \frac{\pi^{2}}{3}{+\ln}%
^{2}{{\frac{Q^{2}}{\mu^{2}}}}\right)  \right]  \left(  \frac{\alpha_{s}%
(\mu^{2})}{\pi}\right)  ^{3}{+}\mathcal{O}{(\alpha}_{s}^{4}{)}\right\}
\]
where
\[
\beta_{0}{=}\frac{1}{4}\left(  11{-}\frac{2}{3}n_{f}\right)  {\,,\qquad}%
\beta_{1}{=}\frac{1}{8}\left(  51{-}\frac{19}{3}{n}_{f}\right)  {\,,}%
\]%
\[
F_{2}{=1.98571-0.115295\,n}_{f}{\,,\qquad}F_{3}{=-6.63694-1.20013\,n}%
_{f}{-0.00518\,n}_{f}^{2}{\,,}%
\]
with $\alpha_{s}(Q^{2})$ being the solution of the equation
\begin{equation}
\frac{\pi}{\beta_{0}\alpha_{s}(Q^{2})}-\frac{\beta_{1}}{\beta_{0}^{2}}%
\ln\left[  \frac{\pi}{\beta_{0}\alpha_{s}(Q^{2})}+\frac{\beta_{1}}{\beta
_{0}^{2}}\right]  =\ln\frac{Q^{2}}{\Lambda^{2}}\,. \label{AlphaRG}%
\end{equation}
In (\ref{Dope}) along with standard power corrections due to the gluon and
quark condensates \cite{SVZII79} we include the unconventional term suppressed
as, $\sim1/Q^{2}$. Its appearance was augmented in \cite{ChetNar} and also
found in the ILM \cite{DoBr03}.

\begin{figure}[h]
\includegraphics[height=7.5cm]{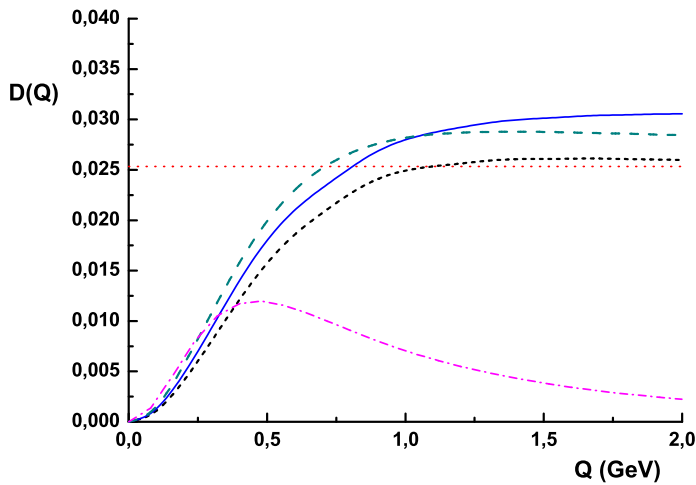}\caption{The Adler function from
the ILM contributions: dynamical quark loop (short dashed), quark + chiral
loops + vector mesons (full line) versus the ALEPH data (dashed). The
dash-dotted line is the prediction of the constituent quark model (extended NJL) and
the dotted line is the asymptotic freedom prediction, $1/4\pi^{2}$.}%
\label{AdlVfig}%
\end{figure}

In the low-$Q^{2}$ limit it is only rigorously known from the theory that
\begin{equation}
{D}_{V}{(Q}^{{2}}{\rightarrow0)=Q}^{{2}}{D}_{V}^{\prime}{(0)+}\mathcal{O}%
{(Q}^{4}{).} \label{D(0)}%
\end{equation}
It is clear (see also Fig. \ref{AdlVfig}) that the Adler function is very
sensitive to transition between asymptotically free (almost massless current
quarks) region described by (\ref{Dope}), (\ref{DpQCD}) to the hadronic regime
with almost constant constituent quarks where one has (\ref{D(0)}).

\begin{figure}[h]
\begin{center}
\includegraphics[height=6cm]{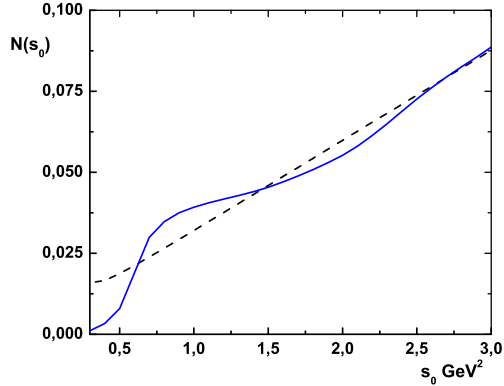}
\end{center}
\caption{The integral, Eq. (\ref{Matching}), versus the upper integration
limit, $s_{0}$, for the $V$ spectral density. The integral of the experimental
data corresponds to solid line and the pQCD prediction (\ref{MatchQCD}) is
given by the dashed line.}%
\label{NormV}%
\end{figure}

To extract the Adler function from experimental data supplemented by QCD
asymptotics (\ref{Dope}), (\ref{DpQCD}) we take following \cite{PPdR98} an
ansatz for the hadronic spectral functions in the form
\begin{equation}
\rho_{J}\left(  t\right)  =\rho_{\mathrm{J}}^{\mathrm{ALEPH}}\left(  t\right)
\theta(s_{0}-t)+\rho_{J}^{\mathrm{pQCD}}\left(  t\right)  \theta(t-s_{0})\,\,,
\label{SpecDens}%
\end{equation}
where
\begin{equation}
\frac{1}{4\pi^{2}}\rho_{V}^{\mathrm{pQCD}}\left(  t\right)  =D_{V}%
^{\mathrm{pQCD}}\left(  t\right)  -\frac{121\pi^{2}}{48}\left(  \frac
{\alpha_{s}(t)}{\pi}\right)  ^{3}, \label{SpecDensQCD}%
\end{equation}
and find the value of continuum threshold $s_{0}$ from the global duality
interval condition:
\begin{equation}
\int_{0}^{s_{0}}dt\,\rho_{\mathrm{J}}^{\mathrm{ALEPH}}(t)=\int_{0}^{s_{0}%
}dt\rho_{J}^{\mathrm{pQCD}}\,(t)\,. \label{Matching}%
\end{equation}
Using the experimental input corresponding to the $\tau$--decay data and the
pQCD expression
\begin{align}
\frac{1}{4\pi^{2}}\int_{0}^{s_{0}}dt\,\rho_{V}^{\mathrm{pQCD}}(t)  &
=\frac{N_{c}}{12\pi^{2}}s_{0}\left\{  1+\frac{\alpha_{\mathrm{s}}(s_{0})}{\pi
}+\left[  F_{2}+\beta_{0}\right]  \left(  \frac{\alpha_{\mathrm{s}}(s_{0}%
)}{\pi}\right)  ^{2}+\right. \label{MatchQCD}\\
&  \left.  +\left[  F_{3}+\left(  2F_{2}\beta_{0}+\beta_{1}\right)
+2\beta_{0}^{2}\right]  \left(  \frac{\alpha_{\mathrm{s}}(s_{0})}{\pi}\right)
^{3}\right\}  ,\nonumber
\end{align}
one finds (Fig. \ref{NormV}) that matching between the experimental data and
theoretical prediction occurs approximately at scale $s_{0}\approx
2.5\mathrm{GeV}^{2}$.

The vector Adler function (\ref{AdlerV}) obtained from matching the low
momenta experimental data and high momenta pQCD asymptotics by using the
spectral density (\ref{SpecDens}) is shown in Fig. \ref{AdlVfig}, where we use
the pQCD asymptotics (\ref{SpecDensQCD}) of the massless vector spectral
function to four loops with $\Lambda_{\overline{\mathrm{MS}}}^{n_{f}=3}=372$
MeV and choose the matching parameter as $s_{0}=2.5$ GeV$^{-1}.$ Admittedly,
in the Euclidean presentation of the data the detailed resonance structure
corresponding to the $\rho$ and $a_{1}$ mesons seen in the Minkowski region
(Fig. \ref{FAleph}) is smoothed out, hence the verification of the theory is
not as stringent as would be directly in the Minkowski space.

The phenomenological definition of the Adler function can be used
for evaluation of the LO contribution to AMM. Below we are going to discuss the
QCD model based definition of the Adler function within the instanton liquid
model \cite{ShSh}. Next two sections we devote to the formulation of the
gauged instanton liquid model \cite{ADoLT00}.

\section{The instanton effective quark model}

Hadronic corrections to AMM are represented as the convolution integrals of
some known kinematical functions times the amplitudes involving low energy
quark processes. To study nonperturbative effects of these amplitudes at low
momenta one can use the framework of the effective field model of QCD. In the
low momenta domain the effect of the nonperturbative structure of QCD vacuum
become dominant. Since invention of the QCD sum rule method based on the use
of the standard OPE it is common to parameterize the nonperturbative
properties of the QCD vacuum by using infinite towers of the vacuum
expectation values of the quark-gluon operators. From this point of view the
nonlocal properties of the QCD vacuum result from the partial resummation of
the infinite series of power corrections, related to vacuum averages of
quark-gluon operators with growing dimension, and may be conventionally
described in terms of the nonlocal vacuum condensates \cite{MikhRad92,DEM97}.
This reconstruction leads effectively to nonlocal modifications of the
propagators and effective vertices of the quark and gluon fields at small momenta.

The adequate model describing this general picture is the instanton liquid
model of QCD vacuum describing nonperturbative nonlocal interactions in terms
of the effective action \cite{ShSh}. Spontaneous breaking of the chiral
symmetry and dynamical generation of a momentum-dependent quark mass are
naturally explained within the instanton liquid model. The nonsinglet and
singlet $V$ and $A$ current-current correlators and the vector Adler function
have been calculated in \cite{DoBr03,ADpepanTop,ADprdG2} in the framework of
the effective chiral model with instanton-like nonlocal quark-quark
interaction \cite{ADoLT00}. In the same model the pion structure function
\cite{DoLT98} and the pion transition form factor normalized by axial anomaly
has been considered in \cite{AD02} for arbitrary photon virtualities. The
nonperturbative properties of the triangle diagram has been thoroughly
discussed in \cite{AD05 WLT,AD05wlts}.

We start with the nonlocal chirally invariant action which describes the
interaction of soft quark fields \cite{ADoLT00}
\begin{align}
&  S&=\int d^{4}x\ \overline{q}_{I}(x)\left[  i\gamma^{\mu}D_{\mu}%
-m_{f}\right]  q_{I}(x)+
\frac{1}{2}G_{P}\int d^{4}X\int\prod_{n=1}^{4}d^{4}x_{n}f(x_{n})
\label{Lint}\\
&&\cdot\left[\overline{Q}(X-x_{1},X)
    \Gamma_{P}Q(X,X+x_{3})\overline{Q}(X-x_{2},X)\Gamma
_{P}Q(X,X+x_{4})\right]  ,\nonumber
\end{align}
where $D_{\mu}=\partial_{\mu}-iV_{\mu}\left(  x\right)  -i\gamma_{5}A_{\mu
}\left(  x\right)  $ and the spin-flavor structure of the nonlocal chirally
invariant interaction of soft quarks is given by the matrix
products\footnote{The explicit calculations below are performed in $SU_{f}(2)$
sector of the model.}
\begin{equation}
G_P (\Gamma_{P}\otimes\Gamma_{P}):\quad G\left(  1\otimes1+i\gamma_{5}\tau
^{a}\otimes i\gamma_{5}\tau^{a}\right)  ,\qquad G^{\prime}\left(  \tau
^{a}\otimes\tau^{a}+i\gamma_{5}\otimes i\gamma_{5}\right)  ,
\end{equation}
where $G$ and $G^{\prime}$ are the 4-quark couplings in the iso-triplet and
iso-singlet channels, and $\tau^{a}$ are the Pauli isospin matrices. For the
interaction in the form of 't Hooft determinant one has the relation
$G^{\prime}=-G$. In general due to repulsion in the singlet channel the
relation $G^{\prime}<G$ is required. In Eq.~(\ref{Lint}) $\overline{q}%
_{I}=(\overline{u},\overline{d})$ denotes the flavor doublet field of
dynamically generated quarks. The separable nonlocal kernel of the interaction
determined in terms of form factors $f(x)$ is motivated by instanton model of
QCD vacuum.

In order to make the nonlocal action gauge-invariant with respect to external
gauge fields $V_{\mu}^{a}(x)$ and $A_{\mu}^{a}(x)$, we define in (\ref{Lint})
the delocalized quark field, $Q(x),$ by using the Schwinger gauge phase
factor
\begin{align}
&  Q(x,y)=P\exp\left\{  i\int_{x}^{y}dz_{\mu}\left[  V_{\mu}^{a}(z)+\gamma
_{5}A_{\mu}^{a}(z)\right]  T^{a}\right\}  q_{I}(y),\nonumber\\
&  \overline{Q}(x,y)=Q^{\dagger}(x,y)\gamma^{0}, \label{Qxy}%
\end{align}
where $P$ is the operator of ordering along the integration path, with $y$
denoting the position of the quark and $x$ being an arbitrary reference point.
The conserved vector and axial-vector currents have been derived earlier in
\cite{ADoLT00,DoBr03,ADprdG2}.

The dressed quark propagator, $S(p)$, is defined as
\begin{equation}
S^{-1}(p)=i\widehat{p}-M(p^{2}), \label{QuarkProp}%
\end{equation}
with the momentum-dependent quark mass found as the solution of the gap
equation
\begin{equation}
M(p^{2})=m_{f}+4G_{P}N_{f}N_{c}f^{2}(p^{2})\int\frac{d^{4}k}{\left(
2\pi\right)  ^{4}}f^{2}(k^{2})\frac{M(k^{2})}{k^{2}+M^{2}(k^{2})}.
\label{SDEq}%
\end{equation}
The formal solution is expressed as \cite{Birse98}
\begin{equation}
M(p^{2})=m_{f}+(M_{q}-m_{f})f^{2}(p^{2}),
\end{equation}
with constant $M_{q}\equiv M(0)$ determined dynamically from Eq.~(\ref{SDEq})
and the momentum dependent $f(p)$ is the normalized four-dimensional Fourier
transform of $f(x)$ given in the coordinate representation.

The nonlocal function $f(p)$ describes the momentum distribution of quarks in
the nonperturbative vacuum. Given nonlocality $f(p)$ the light quark
condensate in the chiral limit, $M(p)=M_{q}f^{2}(p)$, is expressed as
\begin{equation}
\left\langle 0\left\vert \overline{q}q\right\vert 0\right\rangle =-N_{c}%
\int\frac{d^{4}p}{4\pi^{4}}\frac{M(p^{2})}{p^{2}+M^{2}(p^{2})}. \label{QQI}%
\end{equation}
Its $n$-moment is proportional to the vacuum expectation value of the quark
condensate with the covariant with respect to gluon field derivative squared
$D^{2}$ to the $n$th power
\begin{equation}
\left\langle 0\left\vert \overline{q}D^{2n}q\right\vert 0\right\rangle
=-N_{c}\int\frac{d^{4}p}{4\pi^{4}}p^{2n}\frac{M(p^{2})}{p^{2}+M^{2}(p^{2})}.
\label{20}%
\end{equation}
The $n$th moment of the quark condensate appears as a coefficient of Taylor
expansion of the nonlocal quark condensate defined as \cite{MikhRad92}
\begin{equation}
C(x)=\left\langle 0\left\vert \overline{q}\left(  0\right)  P\exp\left[
i\int_{0}^{x}A_{\mu}\left(  z\right)  dz_{\mu}\right]  q\left(  x\right)
\right\vert 0\right\rangle \label{NLC}%
\end{equation}
with gluon Schwinger phase factor inserted for gauge invariance and the
integral is over the straight line path. Smoothness of $C(x)$ near $x^{2}=0$
leads to existence of the quark condensate moments in the \textit{l.h.s.} of
(\ref{20}) for any $n$. In order to make the integral in the \textit{r.h.s.}
of (\ref{20}) convergent the nonlocal function $f(p)$ for large arguments must
decrease faster than any inverse power of $p^{2}$, \textit{e.g.}, like some
exponential
\begin{equation}
f\left(  p\right)  \sim\exp\left(  -\mathrm{const\cdot}p^{\alpha}\right)
,~\alpha>0\quad\mathrm{as\quad}p^{2}\rightarrow\infty. \label{ExpSup}%
\end{equation}

Note, that the operators entering the matrix elements in (\ref{20}) and
(\ref{NLC}) are constructed from the QCD quark and gluon fields. The
\textit{r.h.s.} of (\ref{20}) is the value of the matrix elements of
QCD\ defined operators calculated within the effective instanton model with
dynamical quark fields. Within the instanton model the zero mode function
$f(p)$ depends on the gauge. It is implied \cite{DEM97,DoLT98} that the
\textit{r.h.s.} of (\ref{20}) corresponds to calculations in the axial gauge
for the quark effective field. It is selected among other gauges because in
this gauge the covariant derivatives become ordinary ones: $D\rightarrow
\partial,$ and the exponential in (\ref{NLC}) with straight line path is
reduced to unit. In particular it means that one uses the quark zero modes in
the instanton field given in the axial gauge when define the gauge dependent
dynamical quark mass. The axial gauge at large momenta has exponentially
decreasing behavior and all moments of the quark condensate exist. In
principle, to calculate the gauge invariant matrix element corresponding to
the of \textit{l.h.s.} of (\ref{20}) it is possible to use the expression for
the dynamical mass given in any gauge, but in that case the factor $p^{2n}$
will be modified by more complicated weight function providing invariance of
the answer\footnote{If one would naively use the dynamical quark mass
corresponding to popular singular gauge then one finds the problem with
convergence of the integrals in (\ref{20}), because in this gauge there is
only powerlike asymptotics of $M\left(  p\right)  \sim p^{-6}$ at large
$p^{2}.$}.

Furthermore, the large distance asymptotics of the instanton solution is also
modified by screening effects due to interaction of instanton field with
surrounding physical vacuum \cite{DEM97,ADWB01}. To take into account these
effects and make numerics simpler we shell use for the nonlocal function the
Gaussian form
\begin{equation}
f(p)=\exp\left(  -p^{2}/\Lambda^{2}\right)  , \label{MassDyna}%
\end{equation}
where the parameter $\Lambda$ characterizes the nonlocality size of gluon
vacuum fluctuations and it is proportional to the inverse average size of
instanton in the QCD\ vacuum.

The important property of the dynamical mass (\ref{SDEq}) is that at low
virtualities its value is close to the constituent mass, while at large
virtualities it goes to the current mass value. As we will see below this
property is crucial in obtaining the anomaly at large momentum transfer. The
instanton liquid model can be viewed as an approximation of large-$N_{c}$ QCD
where the only new interaction terms, retained after integration of the high
frequency modes of the quark and gluon fields down to a nonlocality scale
$\Lambda$ at which spontaneous chiral symmetry breaking occurs, are those
which can be cast in the form of four-fermion ope\-ra\-tors (\ref{Lint}). The
parameters of the model are then the nonlocality scale $\Lambda$ and the
four-fermion coupling constant $G_{P}$.

\section{Conserved vector and axial-vector currents}

\begin{figure}[ptb]
\par
\begin{center}
\resizebox{0.75\textwidth}{!}{  \includegraphics{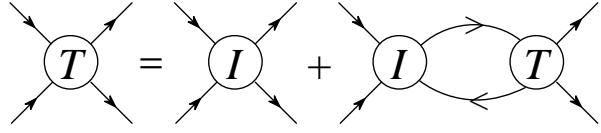}
}
\end{center}
\par
\caption{Diagrammatic representation of the Bethe-Salpeter equation for the
quark-quark scattering matrix, $T$, with nonlocal instanton kernel, $I$.}%
\label{BSTfig}%
\end{figure}

The quark-antiquark scattering matrix (Fig. \ref{BSTfig}) in pseudoscalar
channel is found from the Bethe-Salpeter equation as
\begin{equation}
\widehat{T}_{P}(q^{2})=\frac{G_{P}}{1-G_{P}J_{PP}(q^{2})}, \label{ScattMatr}%
\end{equation}
with the polarization operator being
\begin{equation}
J_{PP}(q^{2})=\int\frac{d^{4}k}{\left(  2\pi\right)  ^{4}}f^{2}\left(
k\right)  f^{2}\left(  k+q\right)  Tr\left[  S(k)\gamma_{5}S\left(
k+q\right)  \gamma_{5}\right]  . \label{J}%
\end{equation}

The position of pion state is determined as the pole of the scattering matrix
\begin{equation}
\left.  \det(1-G_{P}J_{PP}(q^{2}))\right\vert _{q^{2}=-m_{\pi}^{2}}=0.
\label{PoleEq}%
\end{equation}
The quark-pion vertex found from the residue of the scattering matrix is
$\left(  k^{\prime}=k+q\right)  $
\begin{equation}
\Gamma_{\pi}^{a}\left(  k,k^{\prime}\right)  =g_{\pi qq}i\gamma_{5}%
f(k)f(k^{\prime})\tau^{a}\quad
\end{equation}
with the quark-pion coupling found from
\begin{equation}
g_{\pi q}^{-2}=-\left.  \frac{dJ_{PP}\left(  q^{2}\right)  }{dq^{2}%
}\right\vert _{q^{2}=-m_{\pi}^{2}}, \label{gM}%
\end{equation}
where $m_{\mathrm{\pi}}$ is physical mass of the $\pi$-meson. The quark-pion
coupling, $g_{\pi q}$, and the pion decay constant, $f_{\pi}$, are connected
by the Goldberger-Treiman relation, $g_{\pi}=M_{q}/f_{\pi},$ which is verified
to be valid in the nonlocal model \cite{Birse98}, as requested by the chiral symmetry.

\begin{figure}[ptb]
\par
\begin{center}
\resizebox{0.75\textwidth}{!}{  \includegraphics{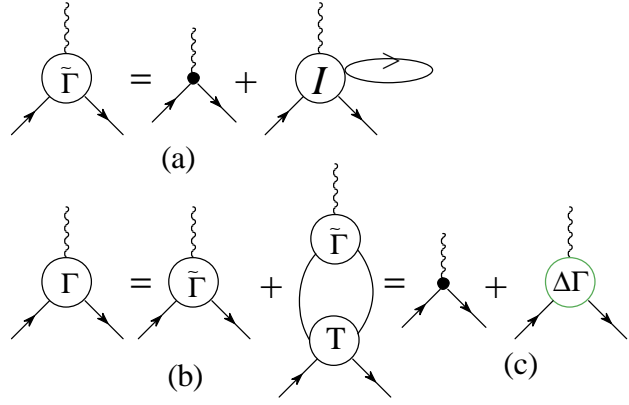}
}
\end{center}
\par
\caption{Diagrammatic representation of the bare (a) and full (b)
quark-current vertices. Diagram (c) shows separation of local (fat dot) and
nonlocal parts of the full vertex.}%
\label{w5}%
\end{figure}

The vector vertex following from the model (\ref{Lint}) is (Fig. \ref{w5}a)
\begin{equation}
\Gamma_{\mu}(k,k^{\prime})=\gamma_{\mu}+(k+k^{\prime})_{\mu}M^{(1)}%
(k,k^{\prime}), \label{GV}%
\end{equation}
where $M^{(1)}(k,k^{\prime})$ is the finite-difference derivative of the
dynamical quark mass (see below (\ref{FDD})), $q$ is the momentum corresponding to the current, and
$k$ $(k^{\prime})$ is the incoming (outgoing) momentum of the quark,
$k^{\prime}=k+q$.

The full axial vertex corresponding to the conserved axial-vector current is
obtained after resummation of quark-loop chain that results in appearance of
term proportional to the pion propagator \cite{ADoLT00} (Fig. \ref{w5}b)
\begin{align}
\Gamma_{\mu}^{5}(k,k^{\prime})  &  =\gamma_{\mu}\gamma_{5}+2\gamma_{5}%
\frac{q_{\mu}}{q^{2}}f(k)f(k^{\prime})\left[  J_{AP}\left(  0\right)
-\frac{m_{f}G_{P}J_{P}\left(  q^{2}\right)  }{1-G_{P}J_{PP}\left(
q^{2}\right)  }\right] \nonumber\\
&  +(k+k^{\prime})_{\mu}J_{AP}\left(  0\right)  \frac{\left(  f(k^{\prime
})-f\left(  k\right)  \right)  ^{2}}{k^{\prime2}-k^{2}}, \label{GAtot}%
\end{align}
where we have introduced the notations
\begin{equation}
J_{P}(q^{2})=\int\frac{d^{4}k}{\left(  2\pi\right)  ^{4}}f\left(  k\right)
f\left(  k+q\right)  Tr\left[  S(k)\gamma_{5}S\left(  k+q\right)  \gamma
_{5}\right]  . \label{JP}%
\end{equation}%
\begin{equation}
J_{AP}(q^{2})=4N_{c}N_{f}\int\frac{d^{4}l}{\left(  2\pi\right)  ^{4}}%
\frac{M\left(  l\right)  }{D\left(  l\right)  }\sqrt{M\left(  l+q\right)
M\left(  l\right)  }. \label{JAP}%
\end{equation}
The axial-vector vertex has a pole at
\[
q^{2}=-m_{\pi}^{2}=m_{c}\left\langle \overline{q}q\right\rangle /f_{\pi}^{2}%
\]
where the Goldberger-Treiman relation and definition of the quark condensate
have been used. The pole is related to the denominator $1-G_{P}J_{PP}\left(
q^{2}\right)  $ in Eq. (\ref{GAtot}), while $q^{2}$ in denominator is
compensated by zero from square brackets in the limit $q^{2}\rightarrow0.$
This compensation follows from expansion of $J(q^{2})$ functions near zero
\begin{align}
J_{PP}(q^{2})  &  =G_{P}^{-1}+m_{c}\left\langle \overline{q}q\right\rangle
M_{q}^{-2}-q^{2}g_{\pi q}^{-2}+O\left(  q^{4}\right)  ,\qquad\\
J_{AP}(q^{2}=0)  &  =M_{q},\qquad J_{P}(q^{2}=0)=\left\langle \overline
{q}q\right\rangle M_{q}^{-1}.
\end{align}
In the chiral limit $m_{f}=0$ the second structure in square brackets in Eq.
(\ref{GAtot}) disappears and the pole moves to zero.

Within the chiral quark model \cite{ADoLT00} based on the non-local structure
of instanton vacuum \cite{DEM97} the full singlet axial-vector vertex
including local and nonlocal pieces is given by (in chiral limit) \cite{DoBr03}
\begin{align}
\Gamma_{\mu}^{5\left(  0\right)  }(k,q,k^{\prime}=k+q) &  =\gamma_{\mu}%
\gamma_{5}+\gamma_{5}(k+k^{\prime})_{\mu}M_{q}\frac{\left(  f\left(
k^{\prime}\right)  -f\left(  k\right)  \right)  ^{2}}{k^{\prime2}-k^{2}%
}+\label{G50}\\
&  +\gamma_{5}\frac{q_{\mu}}{q^{2}}2M_{q}f\left(  k^{\prime}\right)  f\left(
k\right)  \frac{G^{\prime}}{G}\frac{1-GJ_{PP}(q^{2})}{1-G^{\prime}J_{PP}%
(q^{2})}.\nonumber\label{Gmm}%
\end{align}
The singlet current (\ref{G50}) does not contain massless pole due to presence
of the $U_{A}\left(  1\right)  $ anomaly. Indeed, as $q^{2}\rightarrow0$ there
is compensation between denominator and numerator in (\ref{G50})
\begin{equation}
\frac{1-GJ_{PP}(q^{2})}{-q^{2}}=G\frac{f_{\pi}^{2}}{M_{q}^{2}}\qquad
\mathrm{as}\quad q^{2}\rightarrow0,\label{NoGold}%
\end{equation}
where $f_{\pi}$ is the pion weak decay constant. In cancellation of the
massless pole the gap equation is used. Instead, the singlet current develops
a pole at the $\eta^{\prime}-$ meson mass\footnote{See footnote 3. Also
we neglect the effect of the axial-pseudoscalar mixing with the longitudinal
component of the flavor singlet $f_{1}$ meson.}
\begin{equation}
1-G^{\prime}J_{PP}(q^{2}=-m_{\eta^{\prime}}^{2})=0,\label{Eta1}%
\end{equation}
thus solving the $U_{A}(1)$ problem. Let us also remind that in the instanton
chiral quark model the connection between the soft gluon and effective quark
degrees of freedom is fixed by the gap equation. In particular, it means that
the four-quark couplings $G(G^{\prime})$ are proportional to the gluon condensate.

The parameters of the model are fixed in a way typical for effective
low-energy quark models. One usually fits the pion decay constant, $f_{\pi}$,
to its experimental value, which in the chiral limit reduces to $86$
\textrm{MeV} \cite{LeutG}. In the instanton model the constant, $f_{\pi}$, is
expressed as
\begin{equation}
f_{\pi}^{2}=\frac{N_{c}}{4\pi^{2}}\int\limits_{0}^{\infty}du\ u\frac
{M^{2}(u)-uM(u)M^{\prime}(u)+u^{2}M^{\prime}(u)^{2}}{D^{2}\left(  u\right)  },
\label{Fpi2_M}%
\end{equation}
where here and below $u=k^{2},$ primes mean derivatives with respect to $u$:
$M^{\prime}(u)=dM(u)/du$, \textit{etc.}, and
\begin{equation}
D\left(  k^{2}\right)  =k^{2}+M^{2}(k).
\end{equation}

One gets the values of the model parameters \cite{ADprdG2}
\begin{equation}
M_{q}=0.24~\mathrm{GeV,}\qquad\Lambda_{P}=1.11~\mathrm{GeV,\quad}%
G_{P}=27.4~\mathrm{GeV}^{-2}. \label{G's}%
\end{equation}
The coupling $G^{\prime}$ is fixed by fitting the meson spectrum.
Approximately one has $G^{\prime}\approx0.1~G$ \cite{Birse98}.

\section{Adler function within ILM.}

\begin{figure}[h]
\begin{center}
\includegraphics[width=11.5cm]{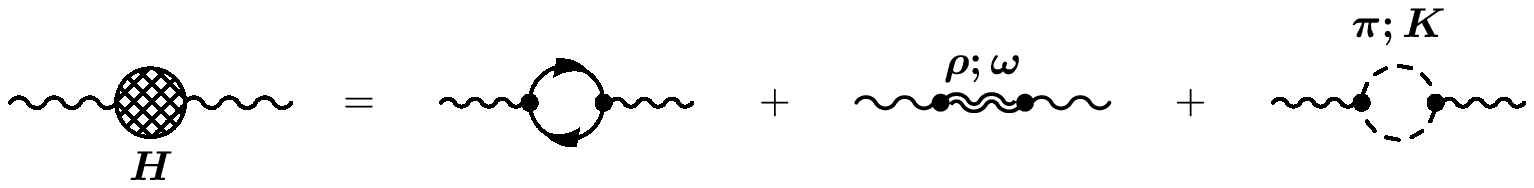}
\end{center}
\caption{Schematic representation of the vector polarization function (\ref{Pncqm}). }%
\label{hpf}%
\end{figure}

Our goal is to obtain the vector current-current correlator and corresponding
Adler function by using the effective instanton-like model (\ref{Lint}) and
then to estimate the leading order hadron vacuum polarization correction to
muon anomalous magnetic moment $a_{\mu}$. In ILM in the chiral limit the
(axial-)vector correlators have transverse character \ \cite{DoBr03}
\begin{equation}
\Pi_{\mu\nu}^{J}\left(  Q^{2}\right)  =\left(  g_{\mu\nu}-\frac{q^{\mu}q^{\nu
}}{q^{2}}\right)  \Pi_{J}^{\mathrm{ILM}}\left(  Q^{2}\right)  , \label{PVmn}%
\end{equation}
where the polarization functions are given by the sum of the dynamical quark
loop, the intermediate (axial-)vector mesons and the higher order mesonic
loops contributions (see Fig. \ref{hpf})
\begin{equation}
\Pi_{J}^{\mathrm{ILM}}\left(  Q^{2}\right)  =\Pi_{J}^{Q\mathrm{Loop}}\left(
Q^{2}\right)  +\Pi_{J}^{\mathrm{mesons}}\left(  Q^{2}\right)  +\Pi_{J}%
^{\chi\mathrm{Loop}}\left(  Q^{2}\right)  . \label{Pncqm}%
\end{equation}

The spectral representation of the polarization function consists
of zero width (axial-)vector resonances $\left(  \Pi_{J}^{\mathrm{mesons}%
}\left(  Q^{2}\right)  \right)  $ and two-meson states $\left(  \Pi_{J}%
^{\chi\mathrm{Loop}}\left(  Q^{2}\right)  \right)  .$ The dynamical quark loop
under condition of analytical confinement has no singularities in physical
space of momenta.

The dominant contribution to the vector current correlator at space-like
momentum transfer is given by the dynamical quark loop which was found in
\cite{DoBr03} with the result\footnote{Within the context of ILM, the
integrals over the momentum are calculated by transforming the integration
variables into the Euclidean space, ($k^{0}\rightarrow ik_{4},$ $k^{2}%
\rightarrow-k^{2}$).}
\begin{align}
\Pi_{V}^{Q\mathrm{Loop}}\left(  Q^{2}\right)   &  =\frac{4N_{c}}{Q^{2}}%
\int\frac{d^{4}k}{\left(  2\pi\right)  ^{4}}\frac{1}{D_{+}D_{-}}\left\{
M_{+}M_{-}+\left[  k_{+}k_{-}-\frac{2}{3}k_{\perp}^{2}\right]  _{ren}\right.
\label{Ploop}\\
&  +\left.  \frac{4}{3}k_{\perp}^{2}\left[  \left(  M^{\left(  1\right)
}\left(  k_{+},k_{-}\right)  \right)  ^{2}\left(  k_{+}k_{-}-M_{+}%
M_{-}\right)  -\left(  M^{2}\left(  k_{+},k_{-}\right)  \right)  ^{\left(
1\right)  }\right]  \right\}  +\nonumber\\
&  +\frac{8N_{c}}{Q^{2}}\int\frac{d^{4}k}{\left(  2\pi\right)  ^{4}}%
\frac{M\left(  k\right)  }{D\left(  k\right)  }\left[  M^{\prime}\left(
k\right)  -\frac{4}{3}k_{\perp}^{2}M^{\left(  2\right)  }\left(
k,k+Q,k\right)  \right]  ,\nonumber
\end{align}
where the notations%

\[k_{\pm}=k\pm Q/2,\qquad k_{\perp}^{2}=k_{+}k_{-}-\frac{\left(  k_{+}q\right)
\left(  k_{-}q\right)  }{q^{2}},
\]%
\[
M_{\pm}=M(k_{\pm}),\ \ \ \ \ D_{\pm}=D(k_{\pm}),
\]
are used. We also introduce the finite-difference derivatives defined for an
arbitrary function $F\left(  k\right)  $ as
\begin{equation}
F^{(1)}(k,k^{\prime})=\frac{F(k^{\prime})-F(k)}{k^{\prime2}-k^{2}},\qquad
F^{(2)}\left(  k,k^{\prime},k^{\prime\prime}\right)  =\frac{F^{(1)}%
(k,k^{\prime\prime})-F^{(1)}(k,k^{\prime})}{k^{\prime\prime2}-k^{\prime2}}.
\label{FDD}%
\end{equation}

\begin{figure}[h]
\begin{center}
\includegraphics[width=11.5cm]{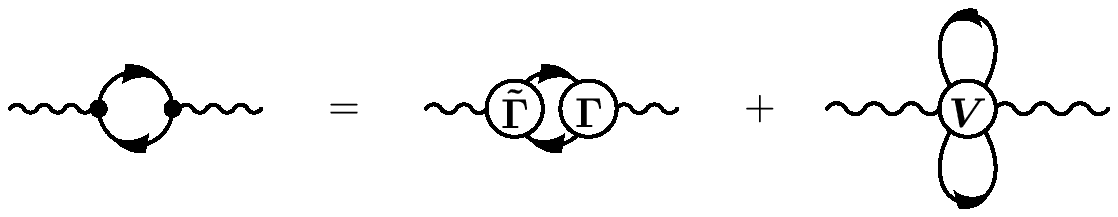}
\end{center}
\caption{The dynamical quark-loop contribution is the sum of dispersive and
contact terms. In the dispersive diagram $\widetilde{\Gamma}$ is the bare
vertex and $\Gamma$ is the total one.}%
\label{DispCont}%
\end{figure}

In (\ref{Ploop}) the first integral represents the contribution of the
dispersive diagrams and the second integral corresponds to the contact
diagrams (see Fig. \ref{DispCont} and ref. \cite{DoBr03} for details). The
expression for $\Pi_{V}^{Q\mathrm{Loop}}\left(  Q^{2}\right)  $ is formally
divergent and needs proper regularization and renormalization procedures which
are symbolically noted by $\left[  ..\right]  _{ren}$ for the divergent term.
At the same time the corresponding Adler function is well defined and finite.

Also we have checked that there is no pole in the vector correlator as
$Q^{2}\rightarrow0$, which simply means that photon remains massless with
inclusion of strong interaction. In the limiting cases the Adler function
derived from Eq. (\ref{Ploop}) in accordance with the first equality of Eq.
(\ref{AdlerV}) satisfies general requirements of QCD (see leading terms in
(\ref{Dope}), (\ref{DpQCD}), and (\ref{D(0)}))%
\begin{equation}
A_{V}^{\mathrm{ILM}}\left(  Q^{2}\rightarrow0\right)  =\mathcal{O}\left(
Q^{2}\right)  ,\qquad A_{V}^{\mathrm{ILM}}\left(  Q^{2}\rightarrow
\infty\right)  =\frac{N_{c}}{12\pi^{2}}+\frac{O_{2}^{V}}{Q^{2}}+\mathcal{O}%
\left(  Q^{-4}\right)  .\label{Aasympt}%
\end{equation}
The leading high $Q^{2}$ asymptotics comes from the $\left[  k_{+}k_{-}%
-\frac{2}{3}k_{\perp}^{2}\right]  _{\mathrm{ren}}$ term in (\ref{Ploop}),
while the subleading asymptotics is driven by "tachionic" term with
coefficient \cite{DoBr03}%
\begin{equation}
O_{2}^{V}=-\frac{N_{c}}{2\pi^{2}}\int_{0}^{\infty}du\frac{uM\left(  u\right)
M^{\prime}\left(  u\right)  }{D\left(  u\right)  }.\label{Tachion}%
\end{equation}
It is possible to integrate Eq. (\ref{Tachion}) in the dilute liquid approximation,
$u>>M^{2}(u)$,
\begin{equation}
O_{2}^{V}\approx\frac{N_{c}}{4\pi^{2}}M_{q}^{2}\approx4.7\cdot10^{-3}%
\quad\mathrm{GeV}^{2},\label{O2}%
\end{equation}
which is close to exact result \cite{DoBr03} and phenomenological estimate
from \cite{ChetNar}.

In the extended by vector interaction model (\ref{Lint}) one gets the
corrections due to the inclusion of $\rho$ and $\omega$ mesons which appear as
a result of quark-antiquark rescattering in these channels%
\begin{equation}
\Pi_{V}^{\mathrm{mesons}}\left(  Q^{2}\right)  =\frac{1}{2Q^{2}}\frac
{G_{V}B_{V}^{2}\left(  Q^{2}\right)  }{1-G_{V}J_{V}^{T}\left(  Q^{2}\right)
}, \label{PVmeson}%
\end{equation}
where $B_{V}\left(  q^{2}\right)  $ is the vector meson contribution to
quark-photon transition form factor%
\begin{align}
B_{V}\left(  Q^{2}\right)   &  =8N_{c}i\int\frac{d^{4}k}{\left(  2\pi\right)
^{4}}\frac{f_{+}^{V}f_{-}^{V}}{D_{+}D_{-}}\left[  M_{+}M_{-}-k_{+}%
k_{-}+\right. \label{BV}\\
&  \left.  +\frac{2}{3}k_{\perp}^{2}\left(  1-M^{2(1)}\left(  k_{+}%
,k_{-}\right)  \right)  -\frac{4}{3}k_{\perp}^{2}\frac{f_{-}f^{(1)}%
(k_{-},k_{+})}{D_{-}}\right]  ,\nonumber
\end{align}
and $J_{V}^{T}\left(  q^{2}\right)  $ is the vector meson polarization
function defined in (\ref{J}) with $\Gamma_{\mu}^{T}=\left(  g_{\mu\nu}%
-q_{\mu}q_{\nu}/q^{2}\right)  \gamma_{\nu}$. As a consequence of the
Ward-Takahashi identity one has $B_{V}\left(  0\right)  =0$ as it should be.

To estimate the $\pi^{+}\pi^{-}$ and $K^{+}K^{-}$ vacuum polarization
insertions (chiral loops corrections) one may use the effective meson vertices
generated by the Lagrangian%
\begin{equation}
-ie\ A_{\mu}\left(  \pi^{+}\overleftrightarrow{\partial}_{\mu}\pi^{-}%
+K^{+}\overleftrightarrow{\partial}_{\mu}K^{-}\right)  \ . \label{EffInter}%
\end{equation}
By using the spectral density calculated from this interaction:
\begin{equation}
\rho_{V}^{\chi loop}\left(  t\right)  ={\frac{1}{12}}\left(  1-{\frac{4m_{\pi
}^{2}}{t}}\right)  ^{3/2}\Theta(t-4m_{\pi}^{2})+\left(  \pi\rightarrow
K\ \right)  , \label{SpectrDensChi}%
\end{equation}
one finds the contribution to the Adler function as%
\begin{equation}
D_{V}^{\chi\mathrm{Loop}}\left(  Q^{2}\right)  =\frac{1}{48\pi^{2}}\left[
a\left(  \frac{Q^{2}}{4m_{\pi}^{2}}\right)  +a\left(  \frac{Q^{2}}{4m_{K}^{2}%
}\right)  \right]  , \label{AChiLoop}%
\end{equation}
where
\begin{equation}
a\left(  t\right)  =\frac{1}{t}\left\{  3+t-\frac{3}{2}\sqrt{\frac{t+1}{t}%
}\left[  \operatorname{arctanh}\left(  \frac{1+2t}{2\sqrt{t\left(  t+1\right)
}}\right)  +i\frac{\pi}{2}\right]  \right\}  . \label{a(t)}%
\end{equation}
The estimate (\ref{AChiLoop}) of the chiral loop corrections corresponds to
the point-like mesons which becomes unreliable at large $t,$ where the meson
form factors has to be taken into account. This contribution corresponds to
the lowest order, $O(p^{4})$, calculations in chiral perturbation theory
($\chi$PT), is non-leading in the formal $1/N_{c}$-expansion and provides
numerically small addition. The higher-loop effects become important at higher momenta.

The resulting Adler function in ILM is given by the sum of above
contributions
\begin{equation}
D_{V}\left(  Q^{2}\right)  =D_{V}^{Q\mathrm{Loop}}\left(  Q^{2}\right)
+D_{V}^{\mathrm{mesons}}\left(  Q^{2}\right)  +D_{V}^{\chi\mathrm{Loop}%
}\left(  Q^{2}\right)  . \label{Dncqm}%
\end{equation}

By using set of parameters found in ILM, \label{G's}, the Adler function in the vector
channel (\ref{Dncqm}) is presented in Fig. \ref{AdlVfig} and the model
estimate for the hadronic vacuum polarization to $a_{\mu}$ given by
(\ref{aAd}) is
\begin{equation}
a_{\mu}^{\mathrm{hvp}~\left(  1\right)  ;\mathrm{ILM}}=623\left(  40\right)
\cdot10^{-10}\ , \label{AMMncqm}%
\end{equation}
where the various contributions to $a_{\mu}^{\mathrm{hvp}~\left(  1\right)
;\mathrm{ILM}}$ are%
\begin{equation}
a_{\mu}^{\mathrm{hvp}~\left(  1\right)  ;\mathrm{Qloop}}=533\cdot
10^{-10},\quad a_{\mu}^{\mathrm{hvp}~\left(  1\right)  ;\mathrm{Vmesons}%
}=13\cdot10^{-10},\quad a_{\mu}^{\mathrm{hvp}~\left(  1\right)  ;\mathrm{\chi
Loop}}=77\cdot10^{-10}%
\end{equation}
and the error in (\ref{AMMncqm}) is due to incomplete knowledge of the higher
order effects in nonchiral corrections. One may conclude, that the agreement
of the instanton model estimate with the phenomenological determinations in
Table 1 is rather good, but model approach unlikely reaches the required by
experiment accuracy. Nevertheless, for the higher order hadronic corrections
we are able essentially reduce the theoretical error by using rather
sophisticated effective quark models. The realistic model calculations are a
crucial issue in consideration of the NLO hadronic contributions. Reproducing
the phenomenological determination of $a_{\mu}^{\mathrm{hvp}~\left(  1\right)
}$, it becomes possible to make reliable estimates of $\Delta a_{\mu
}^{\mathrm{EW}}$ and $a_{\mu}^{\mathrm{h.~L\times L}}.$

With the same model parameters one also gets the estimate for the $\alpha^{2}$
hadronic contribution to the $\tau$-lepton anomalous magnetic moments
\begin{equation}
a_{\tau}^{\mathrm{hvp}~\left(  1\right)  ;\mathrm{ILM}}=3.1\left(  0.2\right)
\cdot10^{-6}\ , \label{AMMtauNCQM}%
\end{equation}
which is in agreement with phenomenological determination
\[
a_{\tau}^{\mathrm{hvp}~\left(  1\right)  ;\mathrm{exp}}=\left\{
\begin{array}
[c]{c}%
3.383\left(  0.111\right)  \cdot10^{-6},\qquad\cite{Jegerlehner96},\\
3.536\left(  0.038\right)  \cdot10^{-6},\qquad\cite{Narison01},
\end{array}
\right.  \
\]
and prediction of the gauged nonlocal quark model \cite{HoldLewM94}%
\[
a_{\tau}^{\mathrm{hvp}~\left(  1\right)  ;G\mathrm{NQM}}=3.2\left(
0.1\right)  \cdot10^{-6}.
\]
Thus, we conclude that the LO hadronic corrections obtained within the ILM are
in reasonable agreement with the latest precise phenomenological numbers.
Next, we are going to use the ILM in order to estimate a subset of $\alpha
^{3}$ hadronic contributions to the muon anomalous magnetic moment, $a_{\mu
}^{\mathrm{hvp}}$

\section{$VA\widetilde{V}$ correlator and NLO corrections to $a_{\mu}$}

\begin{figure}[h]
\begin{center}
\includegraphics[width=10cm]{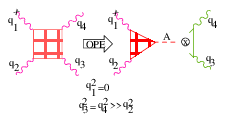}
\end{center}
\caption{OPE\ presentation of the light-by-light scattering as the triangle
amplitude times the coefficient function}%
\label{LBL}%
\end{figure}

Since discovery of anomalous properties \cite{Adler:1969gk,BJ} of the triangle
diagram with incoming two vector and one axial-vector currents
\cite{Rosenberg:1963pp} many new interesting results have been gained.
Recently the interest to triangle diagram has been renewed due to the problem
of accurate calculation of higher order hadronic contributions to muon
anomalous magnetic moment via the light-by-light scattering process (Fig.
\ref{LBL})\footnote{See, \textit{e.g.,} \cite{MelnVain03,CMV,CzarAPP02} and references
therein.}, $a_{\mu}^{\mathrm{h.~L\times L}}$, that cannot be expressed as a
convolution of experimentally accessible observables and need to be estimated
from theory.

\begin{figure}[h]
\begin{center}
\includegraphics[width=4cm]{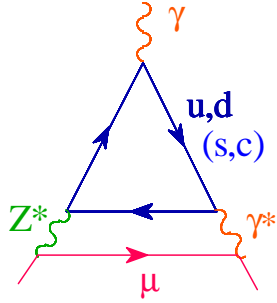}
\end{center}
\caption{Effective $Z^{\ast}\gamma\gamma^{\ast}$ coupling induced by a fermion
triangle contributing to $a_{\mu}^{\mathrm{EW}}$}%
\label{TriZ}%
\end{figure}

The light-by-light scattering amplitude with one photon real and another
photon has the momentum much smaller than the other two, can be analyzed using
operator product expansion (OPE). In this special kinematics the amplitude is
factorized into the amplitude depending on the largest photon momenta and the
triangle amplitude involving the axial current $A$ and two electromagnetic
currents (one soft $\widetilde{V}$ and one virtual $V$). The very similar
kinematics for the triangle amplitude with quark and lepton internal lines
also defines a subset of the two-loop contributions to $a_{\mu}^{\mathrm{EW}}$
via the $Z^{\ast}\gamma\gamma^{\ast}$ effective coupling (Fig. \ref{TriZ})

The corresponding triangle amplitude, which can be viewed as a mixing between
the axial and vector currents in the external electromagnetic field, were
considered recently in \cite{MelnVain03,AD05 WLT,AD05wlts,CMV}. This amplitude can be written as
a correlator of the axial current $j_{\lambda}^{5}$ and two vector currents
$j_{\nu}$ and $\tilde{j}_{\mu}$
\begin{equation}
\widetilde{T}_{\mu\nu\lambda}=-\int\mathrm{d}^{4}x\mathrm{d}^{4}%
y\,\mathrm{e}^{iqx-iky}\,\langle0|\,T\{j_{\nu}(x)\,\tilde{j}_{\mu
}(y)\,j_{\lambda}^{5}(0)\}|0\rangle\,,
\end{equation}
where the currents are defined in (\ref{JAV}), with the tilted current being
for the soft momentum photon vertex. In the specific kinematics when one
photon ($q_{2}\equiv q$) is virtual and another one ($q_{1}$) represents the
external electromagnetic field and can be regarded as a real photon with the
vanishingly small momentum $q_{1}$ depends only on two invariant functions,
longitudinal $w_{L}$ and transversal $w_{T}$ with respect to axial current
index \cite{Kukhto92},%

\begin{align}
&  \widetilde{T}_{\mu\nu\lambda}(q_{1},q)=\frac{1}{4\pi^{2}}\left[
-w_{L}\left(  q^{2}\right)  q^{\lambda}q_{1}^{\rho}q^{\sigma}\varepsilon
_{\rho\mu\sigma\nu}+\right. \nonumber\\
&  \left.  +w_{T}\left(  q^{2}\right)  \left(  q^{2}q_{1}^{\rho}%
\varepsilon_{\rho\mu\nu\lambda}-q^{\nu}q_{1}^{\rho}q_{2}^{\sigma}%
\varepsilon_{\rho\mu\sigma\lambda}+q^{\lambda}q_{1}^{\rho}q_{2}^{\sigma
}\varepsilon_{\rho\mu\sigma\nu}\right)  \right]  . \label{Tt}%
\end{align}
Both structures are transversal with respect to vector current, $q^{\nu
}\widetilde{T}_{\mu\nu\lambda}=0$. As for the axial current, the first
structure is transversal with respect to $q^{\lambda}$ while the second is
longitudinal and thus anomalous.

In the local theory the one-loop result for the invariant functions $w_{T}$
and $w_{L}$ is\footnote{Here and below the small effects of isospin violation
is neglected, considering $m_{f}\equiv m_{u}=m_{d}$.}
\begin{equation}
w_{L}^{\mathrm{1-loop}}=2\,w_{T}^{\mathrm{1-loop}}=\frac{2N_{c}}{3}\int
_{0}^{1}\frac{\mathrm{d}\alpha\,\alpha(1-\alpha)}{\alpha(1-\alpha)q^{2}%
+m_{f}^{2}}\,, \label{wlt}%
\end{equation}
where the factor $N_{c}/3$ is due to color number and electric charge. In the
chiral limit, $m_{f}=0$, one gets the result for space-like momenta $q$
$\left(  q^{2}\geq0\right)  $
\begin{equation}
w_{L}\left(  q^{2}\right)  =2w_{T}\left(  q^{2}\right)  =\frac{2}{q^{2}}.
\label{WLTch}%
\end{equation}

The appearance of the longitudinal structure is the consequence of the axial
Adler-Bell-Jackiw anomaly \cite{Adler:1969gk,BJ}. For the nonsinglet axial
current $A^{\left(  3\right)  }$ there are no perturbative \cite{Adler:er} and
nonperturbative \cite{tHooft} corrections to the axial anomaly and, as
consequence, the invariant function $w_{L}^{\left(  3\right)  }$ remains
intact when interaction with gluons is taken into account. Recently, it was
shown that the relation
\begin{equation}
w_{LT}\left(  q^{2}\right)  \equiv w_{L}\left(  q^{2}\right)  -2w_{T}\left(
q^{2}\right)  =0, \label{wtwl}%
\end{equation}
which holds in the chiral limit at the one-loop level (\ref{WLTch}), gets no
perturbative corrections from gluon exchanges in the iso-singlet case
\cite{VainshPLB03}\footnote{This relation for massive quarks is proved to be valid
up to two-loop level \cite{PasTer05}.}. Nonperturbative nonrenormalization of the nonsinglet
longitudinal part follows from the 't~Hooft consistency condition
\cite{tHooft}, i.e. the exact quark-hadron duality realized as a
correspondence between the infrared singularity of the quark triangle and the
massless pion pole in terms of hadrons. OPE\ analysis indicates that at large
$q$ the leading nonperturbative power corrections to $w_{T}$ can only appear
starting with terms $\sim1/q^{6}$ containing the matrix elements of the
operators of dimension six \cite{Knecht02}. Thus, the transversal part of the
triangle with a soft momentum in one of the vector currents has no
perturbative corrections nevertheless it is modified nonperturbatively.
However, for the singlet axial current $A^{\left(  0\right)  }$ due to the
gluonic $U_{A}\left(  1\right)  $ anomaly there is no massless state even in
the chiral limit. Instead, the massive $\eta^{\prime}$ meson appears. So, one
expects nonperturbative renormalization of the singlet anomalous amplitude
$w_{L}^{\left(  0\right)  }$ at momenta below $\eta^{\prime}$ mass. Below we
demonstrate how the anomalous structure $w_{L}^{\left(  3\right)  }$ is
saturated within the instanton liquid model. We also calculate the transversal
invariant function $w_{T}$ at arbitrary space-like $q$ and show that within
the instanton model in the chiral limit at large $q^{2}$ all allowed by OPE
power corrections to $w_{T}$ cancel each other and only exponentially
suppressed corrections remain \cite{AD05 WLT,AD05wlts}. The nonperturbative
corrections to $w_{T}$ at large $q^{2}$ have exponentially decreasing behavior
related to the short distance properties of the instanton nonlocality in the
QCD vacuum.

The contribution of $Z^{\ast}\gamma\gamma^{\ast}$ vertex to the muon AMM
$a_{\mu}^{\mathrm{EW}}$ in the unitary gauge, where the $Z$ propagator is
$i\left(  -g_{\mu\nu}+q_{\mu}q_{\nu}/m_{Z}^{2}\right)  /\left(  q^{2}%
-m_{Z}^{2}\right)  $, can be written in terms of $w_{L,T}\left(  q^{2}\right)
$ as%
\begin{align}
\Delta a_{\mu}^{\mathrm{EW}} &  =2\sqrt{2}\frac{\alpha}{\pi}G_{\mu}m_{\mu}%
^{2}i\int\frac{d^{4}k}{\left(  2\pi\right)  ^{4}}\frac{1}{q^{2}+2qp}\left[
\frac{1}{3}\left(  1+\frac{2\left(  qp\right)  ^{2}}{q^{2}m_{\mu}^{2}}\right)
\left(  w_{L}-\frac{m_{Z}^{2}}{m_{Z}^{2}-q^{2}}w_{T}\right)  +\right.
\nonumber\\
&  \left.  +\frac{m_{Z}^{2}}{m_{Z}^{2}-q^{2}}w_{T}\right]  ,\label{ammEW}%
\end{align}
where $p$ is the four-momentum of the external muon, $G_{\mu}=1.16637(1)\cdot
10^{-5}$ GeV$^{-2}$ is the Fermi constant obtained from the muon lifetime,
$m_{Z}=91.1875(21)$ GeV, $\alpha=1/137.036$ and for the electron neglecting
its mass one has%
\begin{equation}
w_{L}\left[  e\right]  =2w_{T}\left[  e\right]  =-\frac{2}{Q^{2}}.
\end{equation}
In perturbative QCD with massless quarks the result for the first generation
$\left[  e,u,d\right]  $ contribution is
\begin{equation}
\Delta a_{\mu}^{\mathrm{EW}}\left[  e,u,d\right]  =0,
\end{equation}
due to anomaly cancellation.

\section{$VA\widetilde{V}$ correlator within the instanton liquid model}

\begin{figure}[ptb]
\par
\begin{center}
\resizebox{0.5\textwidth}{!}{  \includegraphics{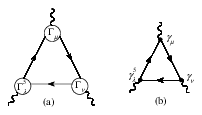}
}
\end{center}
\par
\caption{Diagrammatic representation of the triangle diagram in the instanton
model with dressed quark lines and full quark-current vertices (a); and part
of the diagram when all vertices are local one (b). }%
\label{w6}%
\end{figure}

Our goal is to obtain the nondiagonal correlator of vector current and
nonsinglet axial-vector current in the external electromagnetic field
($VA\widetilde{V}$) by using the effective instanton-like model (\ref{Lint}).
In this model the $VA\widetilde{V}$ correlator is defined by (Fig. \ref{w6}a)
\begin{align}
&  \widetilde{T}_{\mu\nu\lambda}(q_{1},q_{2})=-2N_{c}\int\frac{d^{4}k}{\left(
2\pi\right)  ^{4}}Tr\left[  \Gamma_{\mu}\left(  k+q_{1},k\right)  S\left(
k+q_{1}\right)  \right.  \cdot\nonumber\\
&  \left.  \cdot\Gamma_{\lambda}^{5}\left(  k+q_{1},k-q_{2}\right)  S\left(
k-q_{2}\right)  \Gamma_{\nu}\left(  k,k-q_{2}\right)  S\left(  k\right)
\right]  , \label{Tncqm}%
\end{align}
where the quark propagator, the vector and the axial-vector vertices are given
by (\ref{QuarkProp}), (\ref{GV}) and (\ref{GAtot}), respectively. The
structure of the vector vertices guarantees that the amplitude is transversal
with respect to vector indices
\[
\widetilde{T}_{\mu\nu\lambda}(q_{1},q_{2})q_{1}^{\mu}=\widetilde{T}_{\mu
\nu\lambda}(q_{1},q_{2})q_{2}^{\nu}=0
\]
and the Lorentz structure of the amplitude is given by (\ref{Tt}).

It is convenient to express Eq. (\ref{Tncqm}) as a sum of the contribution
where all vertices are local (Fig. \ref{w6}b), and the rest contribution
containing nonlocal parts of the vertices (Fig. \ref{w6}a). Further results in
this section will concern the chiral limit.

The contributions of diagram \ref{w6}b to the invariant functions at
space-like momentum transfer, $q^{2}\equiv q_{2}^{2}$, are given by
\begin{equation}
w_{L}^{(loc)}\left(  q^{2}\right)  =\frac{4N_{c}}{9q^{2}}\int\frac{d^{4}k}%
{\pi^{2}}\frac{1}{D_{+}^{2}D_{-}}\left[  k^{2}-4\frac{\left(  kq\right)  ^{2}%
}{q^{2}}+3\left(  kq\right)  \right]  , \label{A46a}%
\end{equation}%
\begin{equation}
w_{LT}^{\left(  loc\right)  }\left(  q^{2}\right)  =0, \label{A6a}%
\end{equation}
where we also consider the combination of invariant functions $w_{LT}$, (\ref{wtwl}), which show up
nonperturbative dynamics.
The notations used here and below are
\[
k_{+}=k,\qquad k_{-}=k-q,\qquad k_{\perp}^{2}=k_{+}k_{-}-\frac{\left(
k_{+}q\right)  \left(  k_{-}q\right)  }{q^{2}},
\]%
\begin{equation}
D_{\pm}=D(k_{\pm}^{2}),\qquad M_{\pm}=M(k_{\pm}^{2}),\ \ \ \ f_{\pm}=f(k_{\pm
}^{2}).\ \ \ \
\end{equation}

At large $q^{2}$ one has an expansion
\begin{equation}
w_{L}^{(loc)}\left(  q^{2}\rightarrow\infty\right)  =\frac{2N_{c}}{3}\left(
\frac{1}{q^{2}}+O\left(  q^{-4}\right)  \right)  \label{A4as}%
\end{equation}
It is clear that the contribution (\ref{A46a}) saturate the anomaly at large
$q^{2}$. The reason is that the leading asymptotics of (\ref{A46a}) is given
by the configuration where the large momentum is passing through all quark
lines. Then the dynamical quark mass $M(k)$ reduces to zero and the asymptotic
limit of triangle diagram with dynamical quarks and local vertices coincides
with the standard triangle amplitude with massless quarks and, thus, it is
independent of the model.

The contribution to the form factors when the nonlocal parts of the vector and
axial-vector vertices are taken into account is given by
\begin{align}
w_{L}^{\left(  nonloc\right)  }\left(  q^{2}\right)   &  =\frac{4N_{c}}%
{3q^{2}}\int\frac{d^{4}k}{\pi^{2}}\frac{1}{D_{+}^{2}D_{-}}\left\{
M_{+}\left[  M_{+}-\frac{4}{3}M_{+}^{\prime}k_{\perp}^{2}\right]  -\right.
\nonumber\\
&  \left.  -M^{2(1)}(k_{+},k_{-})\left(  2\frac{\left(  kq\right)  ^{2}}%
{q^{2}}-\left(  kq\right)  \right)  \right\}  . \label{A4bt}%
\end{align}

Summing analytically the local (\ref{A46a}) and nonlocal (\ref{A4bt}) parts
provides us with the result required by the axial anomaly \cite{AD05 WLT}
\begin{equation}
w_{L}(q^{2})=\frac{2N_{c}}{3}\frac{1}{q^{2}}. \label{A4Tot}%
\end{equation}
Fig. \ref{WLfig} illustrates how different contributions saturate the anomaly.
Note, that at zero virtuality the saturation of anomaly follows from anomalous
diagram of pion decay in two photons. This part is due to the triangle diagram
involving nonlocal part of the axial vertex and local parts of the photon
vertices. The result (\ref{A4Tot}) is in agreement with the statement about
absence of nonperturbative corrections to longitudinal invariant function
following from the 't Hooft duality arguments.

For $w_{LT}(q^{2})$ a number of cancellations takes place and the final result
is quite simple \cite{AD05 WLT}%
\begin{align}
&  w_{LT}\left(  q^{2}\right)  =\frac{4N_{c}}{3q^{2}}\int\frac{d^{4}k}{\pi
^{2}}\cdot\nonumber\\
&  \cdot\frac{\sqrt{M_{-}}}{D_{+}^{2}D_{-}}\left\{  \sqrt{M_{-}}\left[
M_{+}-\frac{2}{3}M_{+}^{\prime}\left(  k^{2}+2\frac{\left(  kq\right)  ^{2}%
}{q^{2}}\right)  \right]  -\frac{4}{3}k_{\perp}^{2}\cdot\right. \nonumber\\
&  \left.  \cdot\left[  \sqrt{M_{+}}M^{(1)}(k_{+},k_{-})-2\left(  kq\right)
M_{+}^{\prime}\sqrt{M}^{\left(  1\right)  }(k_{+},k_{-})\right]  \right\}  .
\label{wLT}%
\end{align}

\begin{figure}[ptb]
\par
\begin{center}
\resizebox{0.5\textwidth}{!}{  \includegraphics{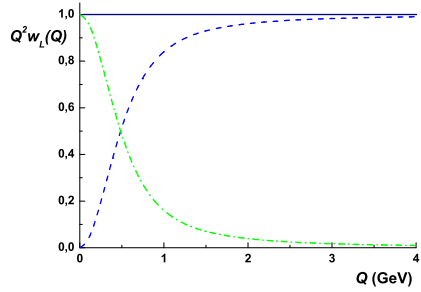}
}
\end{center}
\par
\caption{Normalized $w_{L}$ invariant function constrained by ABJ anomaly from
triangle diagram Fig. \ref{w6}a (solid line) and different contributions to
it: from local part, Fig. \ref{w6}b, (dashed line), and from the nonlocal part
(dash-dotted line).}%
\label{WLfig}%
\end{figure}

\begin{figure}[h]
\hspace*{-1cm} \begin{minipage}{7cm}
\vspace*{0.5cm} \epsfxsize=6cm \epsfysize=5cm \centerline{\epsfbox{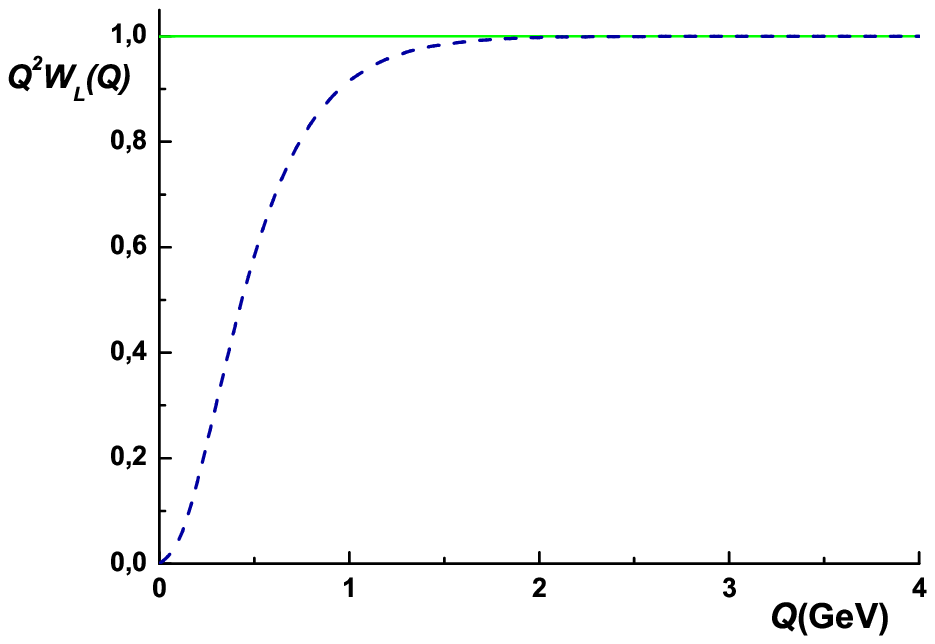}}
\caption[dummy0]{ Normalized $w_L$
invariant function in the nonsinglet case (solid line)
and singlet case (dashed line).
\label{WLfig} }
\end{minipage}\hspace*{0.5cm} \begin{minipage}{7cm}
\vspace*{0.5cm} \epsfxsize=6cm \epsfysize=5cm \centerline{\epsfbox
{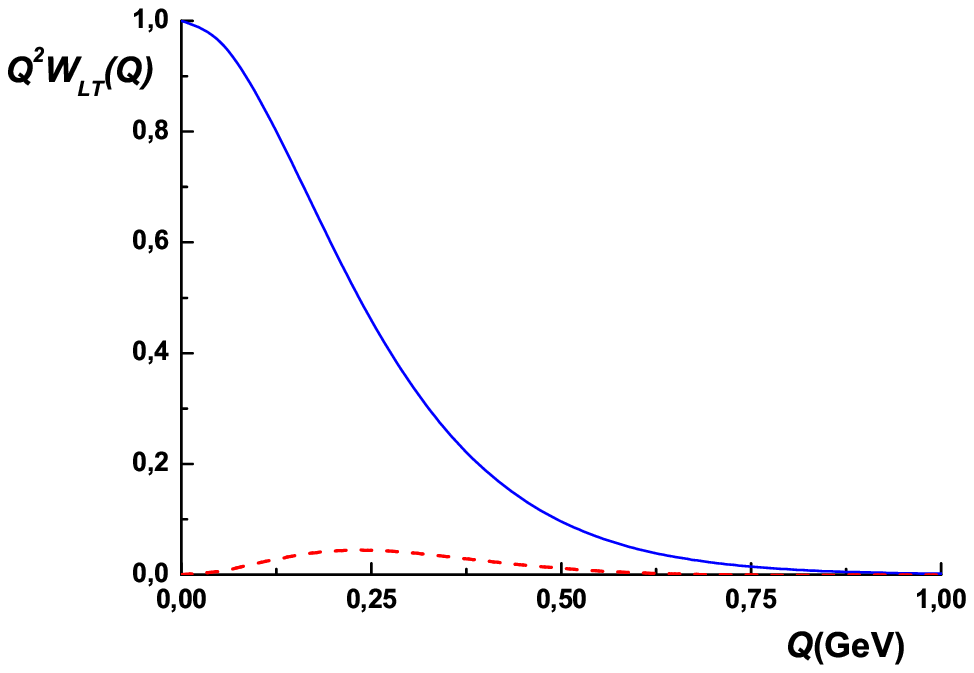}}
\caption[dummy0]{ Normalized $w_{LT}$
invariant function  versus
$Q$ predicted by the instanton model in the nonsinglet case (solid line)
and singlet case (dashed line).
\label{WLTfig} }
\end{minipage}\end{figure}

The behavior of $w_{LT}(q^{2})$ is presented in Fig. \ref{WLTfig}. In the
above expression the integrand is proportional to the product of nonlocal form
factors $f\left(  k_{+}^{2}\right)  f\left(  k_{-}^{2}\right)  $ depending on
quark momenta passing through different quark lines. Then, it becomes evident
that the large $q^{2}$ asymptotics of the integral is governed by the
asymptotics of the nonlocal form factor $f\left(  q^{2}\right)  $ which is
exponentially suppressed (\ref{ExpSup}). Thus, within the instanton model the
distinction between longitudinal and transversal parts is exponentially
suppressed at large $q^{2}$ and all allowed by OPE power corrections are
canceled each other. The instanton liquid model indicates that it may be
possible that due to the anomaly the relation (\ref{wtwl}) is violated at
large $q^{2}$ only exponentially.

The calculations of the singlet $VA\widetilde{V}$ correlator results in the
following modification of the nonsinglet amplitudes \cite{AD05wlts}%
\begin{align}
w_{L}^{\left(  0\right)  }(q^{2})  &  =\frac{5}{3}w_{L}^{\left(  3\right)
}\left(  q^{2}\right)  +\Delta w^{\left(  0\right)  }\left(  q^{2}\right)
,\label{WL0}\\
w_{LT}^{\left(  0\right)  }(q^{2})  &  =\frac{5}{3}w_{LT}^{\left(  3\right)
}\left(  q^{2}\right)  +\Delta w^{\left(  0\right)  }\left(  q^{2}\right)  ,
\label{WLT0}%
\end{align}
where
\begin{align}
\Delta w^{\left(  0\right)  }\left(  q^{2}\right)   &  =-\frac{5N_{c}}{9q^{2}%
}\frac{1-G^{\prime}/G}{1-G^{\prime}J_{PP}\left(  q^{2}\right)  }\int
\frac{d^{4}k}{\pi^{4}}\frac{\sqrt{M_{+}M_{-}}}{D_{+}^{2}D_{-}}\left[
M_{+}-\frac{4}{3}M_{+}^{\prime}k_{\perp}^{2}-\right. \\
&  \left.  -M^{(1)}(k_{+},k_{-})\left(  \frac{4}{3}\frac{\left(  kq\right)
^{2}}{q^{2}}+\frac{2}{3}k^{2}-\left(  kq\right)  \right)  \right]  .
\nonumber\label{DW0}%
\end{align}

Fig. \ref{WLfig} illustrates how the singlet longitudinal amplitude
$w_{L}^{\left(  0\right)  }$ is renormalized at low momenta by the presence of
the $U_{A}\left(  1\right)  $ anomaly. The behavior of $w_{LT}^{\left(
0\right)  }(q^{2})$ is presented in Fig. \ref{WLTfig}. Precise form and even
sign of $w_{LT}^{\left(  0\right)  }(q^{2})$ strongly depend on the ratio of
couplings $G^{\prime}/G$ and has to be defined in the calculations with more
realistic choice of model parameters.

By using (\ref{ammEW}) one finds numerically the result for the first
generation $\left[  e,u,d\right]  $ contribution
\begin{equation}
\Delta a_{\mu}^{\mathrm{EW}}\left[  e,u,d\right]  =-1.48\cdot10^{-11},
\label{AmmEWm}%
\end{equation}
which has to be compared with recent numbers $-2.02\cdot10^{-11}$
\cite{CzMV03} obtained from simple vector dominance model and $-4\cdot
10^{-11}$ \cite{CzKM95} calculated in the naive constituent quark model.

The preliminary estimate of the hadronic light-by-light scattering
contribution within the instanton liquid model is
\begin{equation}
a_{\mu}^{\mathrm{h.~L\times L}}=10.6(1.0)\cdot10^{-10}, \label{AmmLLm}%
\end{equation}
which has to be compared with in $13.6(2.5)\cdot10^{-10}$ \cite{MelnVain03},
where the simple vector meson dominance model has been used.

\section{Conclusions}

We briefly discussed the current status of experimental and theoretical
results on the muon anomalous magnetic moment. The biggest theoretical error
is due to hadronic part of AMM. The phenomenological and model approaches
considered for estimates of leading and next-to-leading order hadronic
corrections to muon AMM. For the model estimates one has used the instanton
liquid model of QCD vacuum. We calculated the vector Adler function and the
nondiagonal correlator of the vector and axial-vector currents in the
background of a soft vector field for arbitrary space-like momenta transfer
and found the corrections to muon anomaly coming from the effects of hadronic
vacuum polarization, $Z^{\ast}\gamma\gamma^{\ast}$ effective vertex and
light-by-light scattering.

The author is grateful to Organizers of the School and in particular to Michal
Praszalowicz for creating of very fruitful atmosphere at the school. The
author also thanks for partial support from the Russian Foundation for Basic
Research projects nos. 03-02-17291, 04-02-16445.

\end{document}